\documentclass[aps,prd,eqsecnum,showpacs,superscriptaddress,twocolumn,nofootinbib,floatfix,preprintnumbers,altaffilletter]{revtex4}
\usepackage{graphicx}
\usepackage{hyperref}
\usepackage{amssymb}\usepackage{amsmath}
\usepackage{longtable}
\usepackage{rotating}
\usepackage{color}
\usepackage{fancyhdr}
\newcommand{\abs}[1]{\left|#1\right|}

\newcommand{\DCCNum}{LIGO-P040006-05-Z}

\def\be{\begin{equation}}
\def\ee{\end{equation}}
\def\ba{\begin{eqnarray}}
\def\ea{\end{eqnarray}}
\def\bi{\begin{itemize}}
\def\ei{\end{itemize}}
\def\ben{\begin{enumerate}}
\def\een{\end{enumerate}}
\def\bw{\begin{widetext}}
\def\ew{\end{widetext}}

\begin{document}

\title{
Optimal combination of signals from co-located gravitational 
wave interferometers\\
for use in searches for a stochastic background}

\author{ Albert Lazzarini}
\affiliation{LIGO Laboratory, California Institute of Technology, 
Pasadena CA 91125, USA}
\author{ Sukanta Bose}
\affiliation{Department of Physics, Washington State University, 
Pullman, WA 99164, USA}
\author{ Peter Fritschel}
\affiliation{LIGO Laboratory, Massachusetts Institute of Technology, 
Cambridge, MA 02139, USA}
\author{ Martin McHugh}
\affiliation{Department of Physics, Loyola University New Orleans, 
New Orleans, LA 70803, USA}
\author{ Tania Regimbau}
\affiliation{Department of Physics \& Astronomy, Cardiff University, 
Cardiff, CF24 3YB, UK}
\author{ Kaice Reilly}
\affiliation{LIGO Laboratory, California Institute of Technology, 
Pasadena CA 91125, USA}
\author{ Joseph D.\ Romano}
\affiliation{Department of Physics \& Astronomy, Cardiff University, 
Cardiff, CF24 3YB, UK}
\author{ John T. Whelan}
\affiliation{Department of Physics, Loyola University New Orleans, 
New Orleans, LA 70803, USA}
\author{ Stan Whitcomb}
\affiliation{LIGO Laboratory, California Institute of Technology, 
Pasadena CA 91125, USA}
\author{ Bernard F.\ Whiting}
\affiliation{Department of Physics, University of Florida, 
Gainesville,  FL 32611, USA}
\date{\today}%

\begin{abstract}
This article derives an {\em optimal\,} (i.e., unbiased, minimum
variance) estimator for the pseudo-detector strain for a pair of
co-located gravitational wave interferometers (such as the pair 
of LIGO interferometers at its Hanford Observatory), allowing for
possible instrumental correlations between the two detectors. 
The technique is robust and does not involve any assumptions or 
approximations regarding the relative strength of
gravitational wave signals in the Hanford pair with respect to other 
sources of correlated instrumental or environmental noise.

An expression is given for the effective power spectral density of
the combined noise in the pseudo-detector. 
This can then be introduced into the standard optimal Wiener 
filter used to cross-correlate detector data streams in order to 
obtain an optimal estimate of the stochastic gravitational wave background. 
In addition, a {\em dual\,} to the optimal estimate of strain is
derived. This dual is constructed to contain {\em no\,} gravitational 
wave signature and can thus be used as an ``off-source" measurement 
to test algorithms used in the ``on-source" observation.
\end{abstract}

\maketitle

\section{Introduction}

Over the past few years a number of long-baseline
interferometric gravitational wave detectors have begun operation.
These include the Laser Interferometer Gravitational
Wave Observatory (LIGO) detectors located in Hanford, WA and
Livingston, LA \cite{ligoproject}; the GEO-600 detector near
Hannover, Germany \cite{GEO-600}; the VIRGO detector near Pisa,
Italy \cite{virgo}; and the Japanese TAMA-300 detector in Tokyo
\cite{tama}. For the foreseeable future all these instruments will
be looking for gravitational wave signals that are expected to be
at the very limits of their sensitivities.
All the collaborations have been developing data analysis
techniques designed to extract weak signals from the detector noise.
Coincidences among multiple detectors will be critical in
establishing the first detections.

In particular, LIGO Laboratory operates two co-located detectors
sharing a common vacuum envelope at its Hanford, WA Observatory
(LHO). One of the two detectors has 4~km long arms and is denoted
H1; the other, with 2~km long arms, is denoted H2. This pair is
unique among all the other kilometer-scale interferometers in the
world because their co-location guarantees {\em simultaneous and
essentially identical\,} responses to gravitational waves. This
fact can provide a powerful discrimination tool for sifting true
signals from detector noise. At the same time, however, the
co-location of the detectors can allow for a greater level of
correlated instrumental noise, complicating the analysis for
gravitational waves.

Indeed, it may not be feasible to ever detect a stochastic gravitational
wave background, or even establish a significant upper limit, via
cross-correlation of H1 and H2, due to the potential of instrumental
correlations. However, even though it may not be profitable to
correlate these co-located detectors, the data from H1 and H2 should
be optimally combined for a correlation analysis with a
geographically separated third detector (such as L1, the LIGO Livingston
detector). 

For the H1-H2 detector pair, properly combining the two data streams
will {\em{always}} result in a pseudo-strain channel that is quieter
than the less noisy detector.  In the limit of completely correlated
noise, this combination could, in principal, lead to a noiseless
estimate of gravitational wave strain. In the other limit where the
detector noise is completely uncorrelated, the two detector outputs can
of course be treated independently and combined at the end of the
analysis to produce a more precise measurement than either separately,
as done in Section V.C. of Ref. \cite{allenromano}. It is the more
general intermediate case, where there is partial correlation of
the detector noise, that is the subject of this paper. 

We show that it is possible to derive an {\em
optimal\,}---i.e., unbiased, minimum variance---strain estimator by
combining the two co-located interferometer outputs into a single,
{\em pseudo-detector\,} estimate of the gravitational wave strain
from the observatory. An expression is given for the effective
power spectral density of the combined noise in the
pseudo-detector. This is then introduced into the standard optimal
Wiener filter used to cross-correlate detector data streams in
order to obtain an estimate of the stochastic gravitational wave
background.

Once the optimal estimator is found, one can subtract this
quantity from the individual interferometer strain channels,
producing a pair of {\em null\,} residual channels for the
gravitational wave signature. The covariance matrix for these two
null channels is Hermitian; it thus possesses two real eigenvalues
and can be diagonalized by a unitary transformation (rotation).
Because the covariance matrix is generated from a single vector,
only one of the eigenvalues is non-zero. The corresponding
eigenvector gives a single null channel that can be used as an
``off-source" channel, which can be processed in the same manner
as the optimal estimator of gravitational wave strain.

The technique described here is possible for the pair of Hanford
detectors because, to high accuracy, the gravitational wave
signature is guaranteed to be {\em identical\,} in both
instruments, and because we can identify specific correlations as
being of instrumental origin. Coherent, time-domain mixing of the
two interferometer strain channels can thus be used to optimal
advantage to provide the best possible estimate of the
gravitational wave strain, and to provide a null channel with
which any gravitational wave analysis can be calibrated for
backgrounds.

The focus of this paper is the development of this technique and
its application to the search for stochastic gravitational waves.
However, it appears that {\em{any}} other search can exploit this
approach.

In Section \ref{sec:II} we discuss the experimental findings during 
recent LIGO science runs which motivated this work to extend the 
optimal filter formalism in the case where instrumental or 
environmental backgrounds are correlated among detectors. 
In Section \ref{sec:III} we introduce the optimal estimate of strain 
for the pair of co-located Hanford interferometers. 
In Section \ref{sec:nogravity} we then introduce the dual null channel. 
Then in Section \ref{sec:V} we apply these formalisms to measurement 
of a stochastic background and consider limiting cases that provide 
insight to understanding the concept. 
Finally in Section \ref{sec:VI} we discuss the implications of 
these results and estimate the effects of imperfect knowledge of 
calibrations on the technique. 
Appendices \ref{sec:general_method}, \ref{sec:application} contain
derivations of formulae used in Sec.~\ref{sec:V}.

\section{Instrumental correlations}
\label{sec:II}

Early operation at LIGO's Hanford observatory has revealed that
the two LHO detectors can exhibit instrumental cross-correlations
of both narrowband and broadband nature. Narrowband correlations
are found, e.g., at the $60$~Hz mains line frequency and
harmonics, and at frequencies corresponding to clocks or timing
signals common in the two detectors; these discrete frequencies
can be identified and removed from the broadband analysis of a
stochastic background search, as described in Ref.
\cite{S1_stochpaper}. Broadband instrumental correlations, on the
other hand, are more pernicious to a stochastic background
analysis; the following types of relatively broadband correlations
have been seen at LHO:
\bi \item Low-frequency seismic excitation of the interferometer
components, up to approximately 15~Hz; at higher frequencies, the
seismic vibrations are not only greatly attenuated by the
detectors' isolation systems, but they also become uncorrelated
over the distances separating the two interferometers. These
correlations are not directly problematic, since they are below
the detection band's lower frequency of $40$~Hz.

\item Acoustic vibrations of the output beam detection systems.

\item Upconversion of seismic noise into the detector band:
intermodulation between the mains line frequencies and the
low-frequency seismic noise produces sidebands around the \{60~Hz,
120~Hz, \ldots\} lines that are correlated between the two
detectors. \ei

Magnetic field coupling to the detectors is another potential
source of correlated noise, though this has not yet been seen to
be significant.

The analysis of the first LIGO science data (S1) for a stochastic
gravitational wave background \cite{S1_stochpaper} showed
substantial cross-correlated noise between the two Hanford
interferometers (H1 and H2), due to the above sources. This
observation led to disregarding the H1-H2 cross-correlation
measurement as an estimate of the stochastic background signal
strength. Two separate upper limits were obtained for the two
transcontinental pairs, L1-H1 and L1-H2 (L1 denotes the 4~km LIGO
interferometer in Livingston, LA). These were not combined because
of the known cross-correlation contaminating the H1-H2 pair.

Here, we show how to take into account such local instrumental
correlations in an optimal fashion by first combining the two
local interferometer strain channels into a single,
pseudo-detector estimate of the gravitational wave strain from the
Hanford site, and then cross-correlating this pseudo-detector
channel with the single Livingston detector output. In doing this,
we obtain a self-consistent utilisation of the three measurements
to obtain a {\em single\,} estimate of the stochastic background
signal strength $\Omega_{\rm gw}$. In order for this to be valid,
the reasonable assumption is made that there are no broadband 
transcontinental (i.e., L1-H1,
L1-H2) instrumental or environmental correlations. 
This has been empirically observed to be the
case for the S1, S2 and S3 science runs when the L1-H1 and L1-H2 
coherences are calculated over long periods of time
(the S1 findings are discussed in \cite{S1_stochpaper}; 
S2 and S3 analyses are still in progress at the time of this writing).

It is important to point out that the technique presented here 
is robust and does not involve any assumptions or approximations 
regarding the relative strength of
gravitational wave signals in the H1-H2 pair with respect to 
other sources of correlated instrumental or environmental noise. 
Since S1, the
sources of environmental correlation between the Hanford pair have
been largely reduced or eliminated. However, as the overall
detector noise is also reduced, smaller cross-correlations become
significant, so it remains important to be able to optimally
exploit the potential sensitivity provided by this unique pair of
co-located detectors.  

\section{Optimal estimate of strain for the two Hanford detectors}
\label{sec:III}

Assume that the detectors H1 and H2 produce data streams
\begin{eqnarray}
s_{H_1}(t) &:=& h(t) + n_{H_1}(t)\,, \\
s_{H_2}(t) &:=& h(t) + n_{H_2}(t)\,,
\end{eqnarray}
respectively, where $h(t)$ is the gravitational wave strain common 
to both the detectors. In the Fourier domain,
\begin{eqnarray}
\widetilde s_{H_1}(f) &=& \widetilde h(f) + \widetilde n_{H_1}(f)\,,
\label{eqn:sh1}\\
\widetilde s_{H_2}(f) &=& \widetilde h(f) + \widetilde n_{H_2}(f)\,,
\label{eqn:sh2}
\end{eqnarray}
where we defined the Fourier transform of a time domain function, $a(t)$,
as $\widetilde a(f):=\int_{-\infty}^\infty dt\, e^{-i2\pi ft}\,a(t)$.
Also assume that the processes generating $h$, $n_{H_1}$, $n_{H_2}$ 
are stochastic with the following statistical properties:
\begin{eqnarray}
\langle\widetilde n_{H_i}(f)\rangle = 
\langle\widetilde h(f)\rangle 
&=& 
0\,,
\\
\langle\widetilde n_{H_i}^*(f)\widetilde h(f)\rangle 
&=&
0\,,
\\
\langle\widetilde n_{H_i}^*(f)\widetilde n_{H_j}(f')\rangle 
&=& 
P^n_{H_i H_j}(f)\,\delta(f-f')\,, \label{eqn:theorycrosspower}
\\
\langle\widetilde h^*(f)\widetilde h(f')\rangle 
&=& 
P_{\Omega}(f)\,\delta(f-f')\,,
\\
\langle\widetilde s_{H_i}^*(f)\widetilde s_{H_j}(f')\rangle&:=& P_{H_i H_j}(f)\,\delta(f-f')\,\label{eqn:measurecrosspower}
\\
&=& (P^n_{H_i H_j}(f)+P_{\Omega}(f))\, \\
&&\times \delta(f-f')
\\
P^n_{H_i H_i}(f) 
&:=&
P^n_{H_i}(f)\,,\label{eqn:theorypower}
\\
P_{H_i H_i}(f) \label{eqn:measurepower}
&:=&
P_{H_i}(f)\,,
\\
\rho_{H_i H_j}(f)
&:=&
\frac{P_{H_i H_j}(f)}{\sqrt{P_{H_i}(f) P_{H_j}(f)}}\,,
\\
\Gamma_{H_i H_j}(f)
&:=&
|\rho_{H_i H_j}(f)|^2\,,
\\
P_{\Omega}(f) 
&\ll& 
P_{H_i}(f)\,,
\label{eqn:pomegaph}
\end{eqnarray}
where $i=1,2$ and the angular 
brackets $\langle...\rangle$ denote ensemble or statistical 
averages of random processes. 
Note that Eqs.~(\ref{eqn:measurecrosspower}) and (\ref{eqn:measurepower})
signify the {\em{measurable}} cross-power and power spectra while 
Eqs.~(\ref{eqn:theorycrosspower}) and (\ref{eqn:theorypower})
refer to intrinsic noise quantities that cannot, in principle, 
be isolated in a measurement. Often, Eq.~(\ref{eqn:pomegaph}) 
is assumed in order to
identify instrument noise power with the measured quantity.
Note also that the coherence $\rho_{H_i H_j}(f)$ is a complex quantity 
of magnitude less than or equal to unity, and that 
$P_{H_j H_i}(f)=P^*_{H_i H_j}(f)$.

Now construct an {\em unbiased\,} linear combination of 
$\widetilde s_{H_i}(f)$:
\begin{equation}
\widetilde s_H(f) := 
\widetilde\alpha(f) \widetilde s_{H_1}(f) + 
(1 - \widetilde\alpha(f)) \widetilde s_{H_2}(f)\,.
\label{eqn:defSh}
\end{equation}
If $\widetilde s_H(f)$ is also to be a {\em minimum variance\,} 
estimator, where
\begin{equation} 
{\rm Var}(s_H) 
:=
\langle\widetilde s^*_H(f) \widetilde s_H(f')\rangle 
= 
P_{H}(f)\,\delta(f-f')\,,
\end{equation}
with
\begin{eqnarray} 
P_{H}(f)
&=&
|\widetilde \alpha(f)|^2 P^n_{H_1}(f) + 
|1-\widetilde \alpha(f)|^2 P^n_{H_2}(f) +
\nonumber
\\
&&+
\widetilde \alpha^*(f) (1-\widetilde \alpha(f))P^n_{H_1 H_2}(f) + 
\nonumber
\\
&&+\widetilde \alpha(f)   (1-\widetilde \alpha^*(f))P^{n*}_{H_1H_2}(f) + 
P_\Omega(f)\,,
\nonumber
\\
\label{eqVars}
\end{eqnarray}
then $\widetilde\alpha(f)$ must have the following form:
\begin{equation}
\widetilde \alpha(f) =
\frac{P_{H_2}(f) - P_{H_1H_2}(f)}
{P_{H_1}(f) + P_{H_2}(f) - (P_{H_1H_2}(f) + P^*_{H_1H_2}(f))}\,.
\label{eqn:alpha1}
\end{equation}
The corresponding power of the pseudo-detector signal is
\begin{equation}
 P_{H}(f)=  
\frac{P_{H_1}(f) P_{H_2}(f) (1-\Gamma_{H_1H_2}(f))}
{P_{H_1}(f) + P_{H_2}(f) - (P_{H_1H_2}(f) + P^*_{H_1H_2}(f))}\,.
\label{eqn:pH1}
\end{equation}

It is important to note that the above
expressions for $\widetilde \alpha(f)$ and $P_H(f)$ do 
{\em not\,} require any assumption on the relative strength
of the cross-correlated stochastic signal to the instrumental
or environmental cross-correlated noise.
In particular, the stochastic signal power $P_\Omega$ enters
$P_{H_1}$, $P_{H_2}$, and $P_{H_1 H_2}$ in exactly the same way,
canceling out in Eq.~(\ref{eqn:alpha1}), implying that the 
above solution for $\widetilde\alpha$ is independent of the 
relative strength of the stochastic signal to other sources of 
cross-correlated noise.
In addition, Eqs.~(\ref{eqn:alpha1}), (\ref{eqn:pH1}) 
involve only {\em experimentally 
measurable\,} power spectra and cross-spectra (and not
the intrinsic noise spectra), indicating that this procedure 
{\em can} be carried out in practice.

\begin{figure*}[htbp!]
\includegraphics[width=6.0in,angle=0]{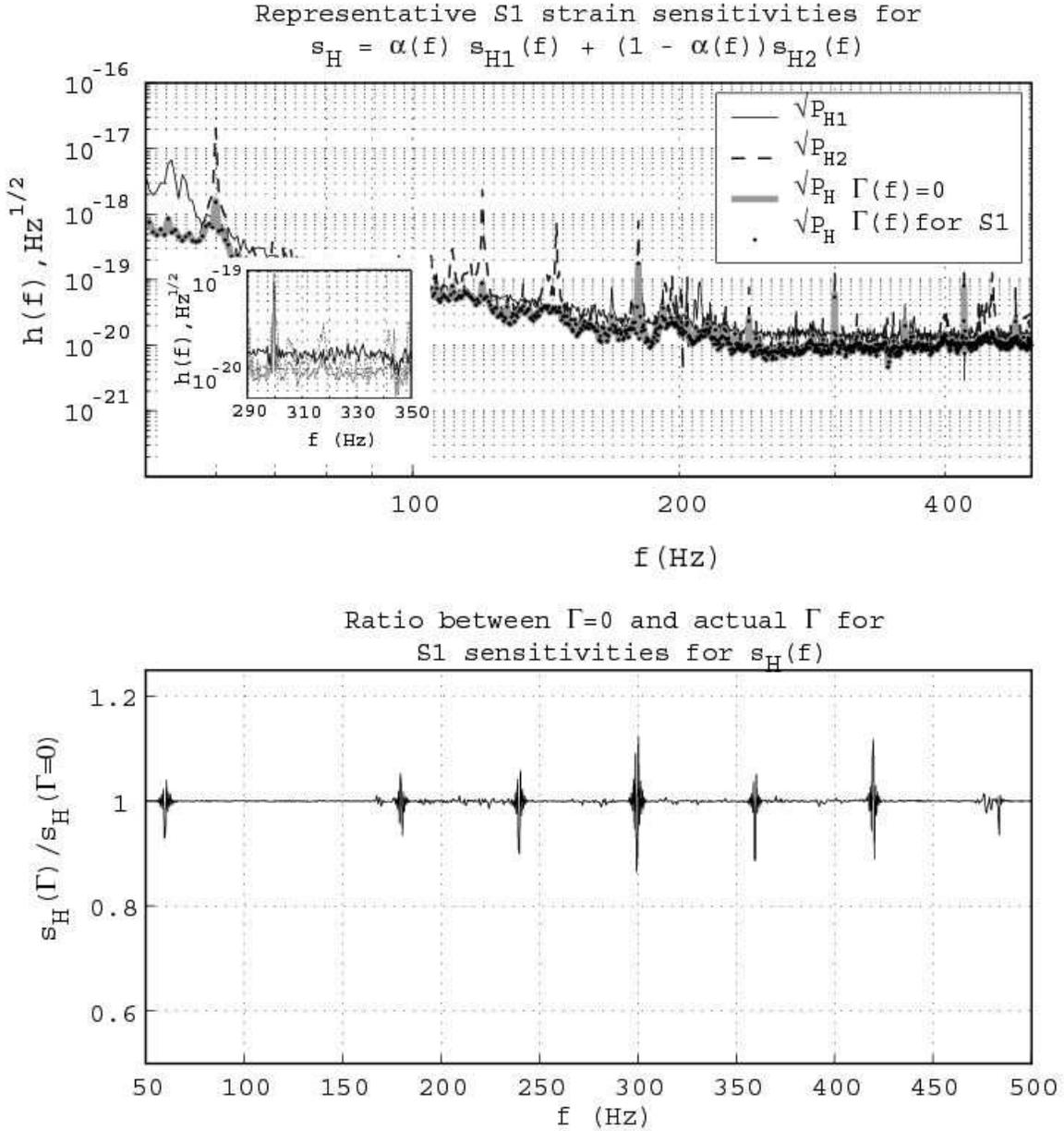}
\caption{Strain spectral densities (i.e., absolute value) of 
$\widetilde s_H(f)$ (gray or dotted), $\widetilde s_{H_1}(f)$ (black), and 
$\widetilde s_{H_2}(f)$ (dashed), representative of the S1 data.
{\bf{Top Panel:}} overlay of the individual spectral 
densities with that of 
the strain spectral density $|\widetilde s_H(f)|$ 
calculated with the S1 run-averaged coherence,  $\Gamma_{H_1 H_2}(f)$, 
and with $\Gamma_{H_1H_2}(f) = 0$. 
On this scale, the left hand panel shows no discernible difference 
between the spectra for $\Gamma_{H_1 H_2}(f)$, and with 
$\Gamma_{H_1H_2}(f) = 0$, 
suggesting that even the level of coherence seen during the S1 run 
might be sufficiently low to allow one to simply
combine the L1-H1 and L2-H2 cross-correlation 
measurements under the assumption of zero cross-correlated
noise. 
The optimality of the estimate $\widetilde s_H(f)$ 
is visible here because it is {\em{always less than the smaller\,}} 
of $\widetilde s_{H_1}(f)$  or $\widetilde s_{H_2}(f)$. The inset shows a blow-up of the region near one of the spectral features. On this scale the individual spectra can be discerned.
{\bf{Bottom panel:}} plot of the ratio of  amplitude spectra for 
$|\widetilde s_H(f)|$ calculated with $\Gamma_{H_1 H_2}(f)$ as 
measured during S1 and $\Gamma_{H_1 H_2}(f)=0$ (i.e., assuming no coherence). 
The difference between the two is very small except for the very 
lowest frequencies and at narrow line features.}
\label{fig:Optimal_sH}
\end{figure*}

\begin{figure*}[htbp!]  
\includegraphics[width=5.5in,angle=0]{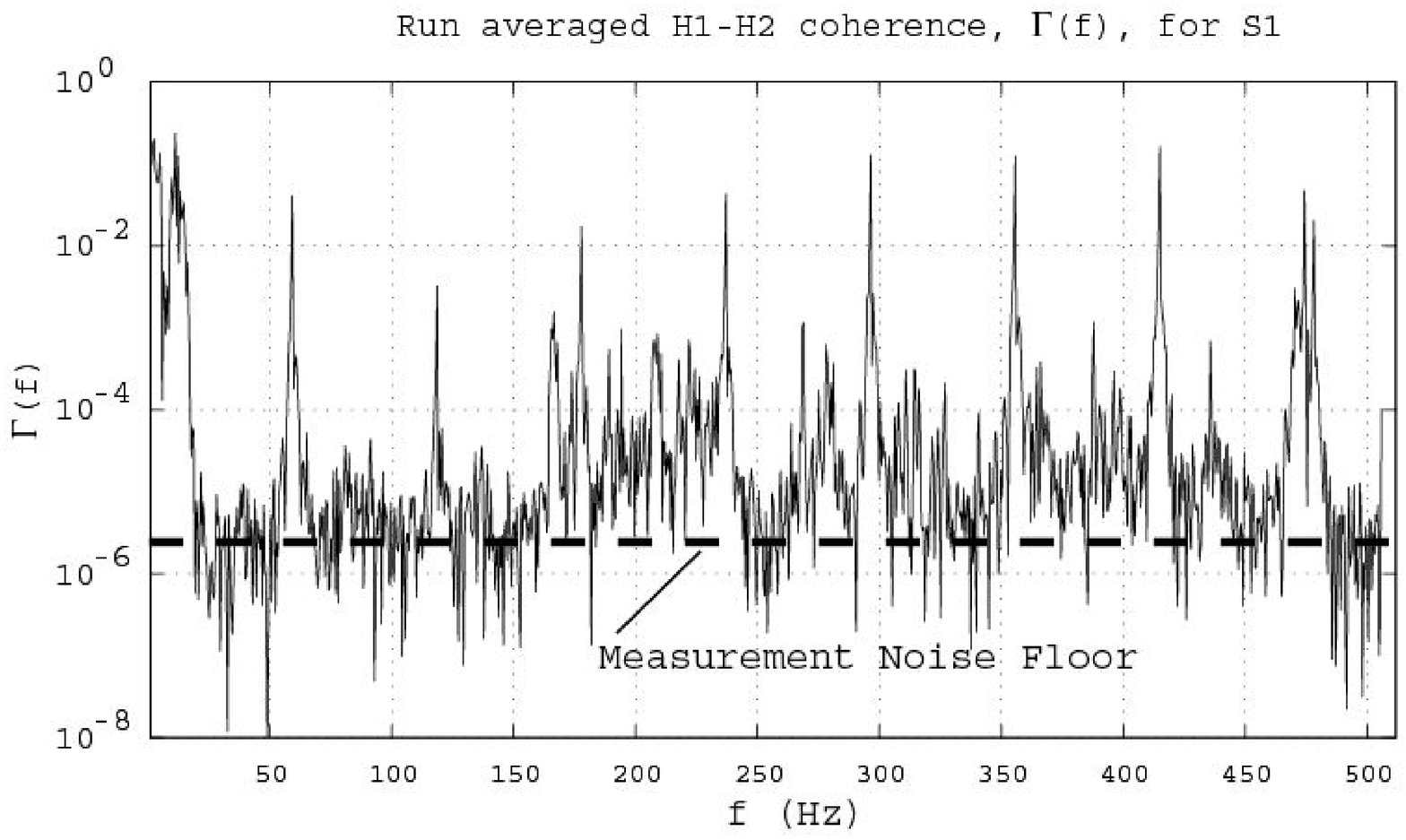}
\caption{H1-H2 coherence averaged over the whole S1 data run.
Note the substantial broadband coherence below 
250~Hz and between 400 and 475~Hz.
Low frequency seismic noise and acoustic coupling between the 
input electro-optics systems are considered to be the prime 
sources of this cross-correlated noise \cite{S1_stochpaper,schofield}.}
\label{fig:GammaH1H2}
\end{figure*}

Figure~\ref{fig:Optimal_sH} shows plots of the strain spectral 
densities for $\widetilde s_H(f)$, 
$\widetilde s_{H_1}(f)$, and $\widetilde s_{H_2}(f)$, 
representative of the S1 data. 
The strain spectral density $|\widetilde s_H(f)|$ is calculated
from Eqs.~(\ref{eqn:defSh}) and (\ref{eqn:alpha1}) for both
$\Gamma_{H_1 H_2}(f)=0$ (i.e., an artificial case that assumes 
no coherence), and for the coherence $\Gamma_{H_1 H_2}(f)$ that 
was actually measured
over the whole S1 data run (see Fig.~\ref{fig:GammaH1H2}).
The plots in Fig.~\ref{fig:Optimal_sH} suggest that the observed 
level of coherence during the S1 run, $\Gamma\sim 10^{-5}$, 
might be sufficiently low that one can simply combine the L1-H1, 
L1-H2 cross-correlation measurements under the assumption of zero 
cross-correlated noise (c.f.\ Eq.~(\ref{e:no_correlations})). 
The formalism developed in this paper allows a quantitative 
assessment of the effect of instrumental or environmental 
correlations on combining independently analyzed results 
{\em{ex post facto}}.

\subsection{Limiting cases}

I. If $\rho_{H_1H_2}(f)=0$, then 
\begin{eqnarray}
\widetilde \alpha(f)
&=&
\frac{P_{H_2}(f)}{P_{H_1}(f)+P_{H_2}(f)}\,,
\label{eqn:alphauncorr}
\\
\widetilde s_H(f)
&=&\frac{P_{H_2}(f)\widetilde s_{H_1}(f) + P_{H_1}(f) \widetilde s_{H_2}(f)}
{P_{H_1}(f) +P_{H_2}(f)}\,,
\label{eqn:uncorrSh}
\\
P_H(f)
&=&
\frac{P_{H_1}(f)P_{H_2}(f)}{P_{H_1}(f) + P_{H_2}(f)}\,.
\label{eqn:uncorrPh}
\end{eqnarray}

II. If $P_{H_1}(f)=P_{H_2}(f)$, then 
\begin{equation}
\widetilde \alpha(f)=
\frac {1-\rho_{H_1H_2}(f)}{2-(\rho_{H_1H_2}(f) + \rho_{H_1H_2}^*(f))}\,.
\end{equation}

III. If $\rho_{H_1H_2}(f)=1$
and $P_{H_1}(f)=P_{H_2}(f)$, then 
\begin{eqnarray}
P_{H}(f)=
\lim_{\Gamma(f)\rightarrow 1}
{\frac{P_{H_1}(f)}{2}}\frac{1-\Gamma(f)}{1-\sqrt{\Gamma(f)}}= 
P_{H_1}(f)\,.
\end{eqnarray}

IV. If $P_{H_2}(f) =  4 P_{H_1}(f)$ (which is the limiting
design performance for H1 and H2 due to the $2:1$ arm length ratio),
then
\begin{equation}
\widetilde \alpha(f)=
\frac
{2 (2-\rho_{H_1H_2}(f))}
{5-2(\rho_{H_1H_2}(f) + \rho_{H_1H_2}^*(f)) }\,.
\end{equation}
Note for this case that if the noise were either completely 
correlated 
($\rho_{H_1H_2}(f)=1 \Rightarrow\widetilde \alpha(f)=2$) 
or anti-correlated 
($\rho_{H_1H_2}(f)=-1\Rightarrow\widetilde \alpha(f)=2/3$), 
then one could exactly cancel the noise from the signals 
$\widetilde s_{H_i}$. 
If the noise is uncorrelated 
($\rho_{H_1H_2}(f)=0\Rightarrow \widetilde \alpha(f)=4/5$), 
then the weighting of the signals from the two interferometers is 
in the ratio $4:1$, as expected.

\section{A dual to the optimal estimate of strain that cancels
the gravitational wave signature} 
\label{sec:nogravity}

In the previous section, an optimal estimator of the gravitational
wave strain $h$ was derived by appropriately combining the 
outputs of the two Hanford detectors. 
It is also possible to form a {\em dual\,} to 
this optimal estimate (denoted $\widetilde z_H(f)$) that 
explicitly cancels the gravitational wave signature. 

\begin{figure*}[H1H2schematic]  
\includegraphics[width=5.5in,angle=0]{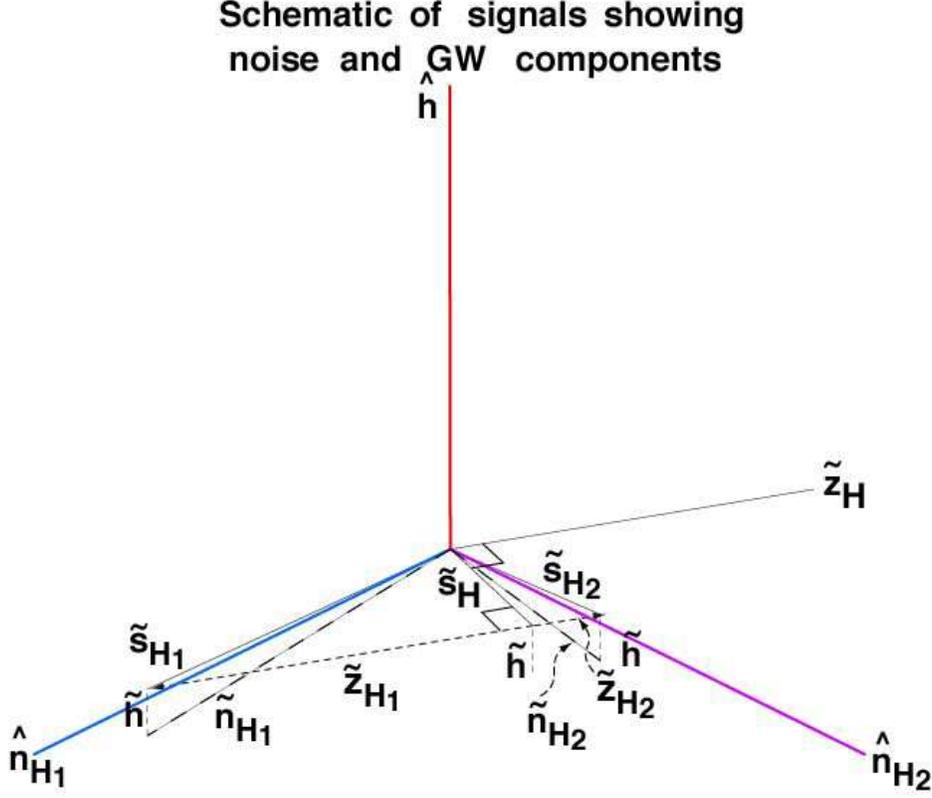}
\caption{Schematic showing how the H1 and H2 signals may be represented 
in a 3-dimensional space of noise components for the two detectors 
and the common gravitational wave strain: 
\{$\hat n_{H_1}, \hat n_{H_2}, \hat h$\}. 
The signals $\widetilde s_{H_1}(f)$ and  $\widetilde s_{H_2}(f)$ are not, 
in general, orthogonal if the coherence between the noise, 
$\widetilde n_{H_1}(f)$ and $\widetilde n_{H_2}(f)$, is non-zero. 
$\widetilde s_{H}(f)$ is the minimum variance estimate of 
$\widetilde h(f)$ derived from $\widetilde s_{H_1}(f)$ and  
$\widetilde s_{H_2}(f)$.  
Using $\widetilde s_{H}(f)$ as the best estimate of $\widetilde h(f)$, 
this signal can be subtracted from $\widetilde s_{H_1}(f)$ and  
$\widetilde s_{H_2}(f)$ to produce the vectors 
$\widetilde z_{H_1}(f)$, $\widetilde z_{H_2}(f)$ that lie in the 
$\hat n_{H_1}$-$\hat n_{H_2}$ plane. 
These vectors give rise to the covariance matrix 
$||\widetilde {\bf{C}}_z(f)||$.  
$\widetilde z_{H_1}(f)$ and $\widetilde z_{H_2}(f)$ are colinear 
and thus one of the eigenvectors of $||\widetilde {\bf{C}}_z(f)||$ 
will be zero. 
The other corresponds to the dual of  $\widetilde s_H(f)$, denoted 
$\widetilde z_H(f)$, which is orthogonal to $\widetilde s_H(f)$, as 
shown in the figure. 
Note that it is necessary to first subtract the contribution of 
$\widetilde h(f)$ from the signals before forming the covariance matrix.}
\label{fig:H1H2schematic}
\end{figure*}

Starting with Eqs.~(\ref{eqn:sh1}), (\ref{eqn:sh2}),  and
the optimal estimate $\widetilde s_H(f)$, we construct the
$h$-subtracted residuals:
\begin{eqnarray} 
\widetilde z_{H_1}(f) 
&:=& 
\widetilde s_{H_1}(f) - \widetilde s_{H}(f)\,, 
\\
\widetilde z_{H_2}(f) 
&:=& 
\widetilde s_{H_2}(f) - \widetilde s_{H}(f)\,.
\end{eqnarray}
Both $\widetilde z_{H_1}(f)$, $\widetilde z_{H_2}(f)$ are 
proportional to $\widetilde n_{H_1}(f)-\widetilde n_{H_2}(f)$,
although with different frequency-dependent weighting functions:
\begin{eqnarray}
\widetilde z_{H_1}(f) 
&=& 
\left(1-\widetilde \alpha(f) \right)\,
\left(\widetilde n_{H_1}(f) -  \widetilde n_{H_2}(f)\right)\,,
\label{eqn:defZh1}
\\
\widetilde z_{H_2}(f) 
&=& -\widetilde \alpha(f)\,
\left(\widetilde n_{H_1}(f) -  \widetilde n_{H_2}(f)\right)\,.
\label{eqn:defZh2}
\end{eqnarray}
Figure~\ref{fig:H1H2schematic} shows schematically the 
geometrical relationships of the signal vectors 
$\widetilde s_{H_i}(f)$ and $\widetilde z_{H_i}(f)$. 
Once the best estimate $\widetilde s_H(f)$ is subtracted 
from the signals, the residuals lie in the 
$\hat n_{H_1}$-$\hat n_{H_2}$ plane.
(Here $\hat n_{H_1}$ and $\hat n_{H_2}$ are unit vectors pointing
in directions corresponding to {\em uncorrelated\,} detector noise.)
Their covariance matrix can then be diagonalized without affecting 
the gravitational wave signature $h$ contained in $\widetilde s_{H}(f)$.

Now consider the covariance matrix 
\begin{equation}
||\widetilde{\bf C}_z||_{ij}\,\delta(f-f'):=
\langle \widetilde z^*_{H_i}(f)\widetilde z_{H_j}(f') \rangle\,.
\end{equation}
Then one can show that
\begin{widetext}
\begin{eqnarray}
||\widetilde{\bf C}_z(f||\,\delta(f-f')
&=&
\left[ 
\begin{array}{c c}
\langle \widetilde z^*_{H_1}(f)\widetilde z_{H_1}(f')\rangle &
\langle \widetilde z^*_{H_1}(f)\widetilde z_{H_2}(f')\rangle
\\
\langle \widetilde z^*_{H_2}(f)\widetilde z_{H_1}(f')\rangle & 
\langle \widetilde z^*_{H_2}(f)\widetilde z_{H_2}(f')\rangle
\end{array} 
\right] 
\\
&=&
\left[ 
\begin{array}{c c}
|1-\widetilde \alpha(f)|^2 &
-\widetilde \alpha(f)+|\widetilde \alpha(f)|^2
\\
-\widetilde \alpha^*(f)+|\widetilde \alpha(f)|^2 & 
|\widetilde \alpha(f)|^2
\end{array} 
\right] 
\langle(\widetilde n^*_{H_1}(f)-\widetilde n^*_{H_2}(f))
       (\widetilde n_{H_1}(f') - \widetilde n_{H_2}(f'))\rangle  
\\
&=&
\left[ 
\begin{array}{c c}
|1-\widetilde \alpha(f)|^2 &
-\widetilde \alpha(f)+|\widetilde \alpha(f)|^2
\\
-\widetilde \alpha^*(f)+|\widetilde \alpha(f)|^2 & 
|\widetilde \alpha(f)|^2
\end{array} 
\right] 
\langle(\widetilde s^*_{H_1}(f)-\widetilde s^*_{H_2}(f))
       (\widetilde s_{H_1}(f') - \widetilde s_{H_2}(f'))\rangle  
\\
&=& 
\left[ 
\begin{array}{c c}
|1-\widetilde \alpha(f)|^2 &
-\widetilde \alpha(f)+|\widetilde \alpha(f)|^2
\\
-\widetilde \alpha^*(f)+|\widetilde \alpha(f)|^2 & 
|\widetilde \alpha(f)|^2
\end{array} 
\right]\times
\nonumber
\\
&&\times
\left(P_{H_1}(f) + P_{H_2}(f) - (P_{H_1H_2}(f) + P^*_{H_1H_2}(f))\right)\,
\delta(f-f')\,.
\label{covCz}
\end{eqnarray}
%
Diagonalization of
$||\widetilde{\bf{C}}_z(f)||$ gives the eigenvalues:
\begin{eqnarray}
\lambda_1&=&0\,,
\\
\lambda_2&=& 
\left(P_{H_1}(f) + P_{H_2}(f) - (P_{H_1H_2}(f) + P^*_{H_1H_2}(f))\right)\,
\left(1-\widetilde\alpha(f) - \widetilde\alpha^*(f) + 
2|\widetilde\alpha(f)|^2\right)\,.
\label{eqn:lambda2}
\end{eqnarray}
The non-trivial solution corresponds to the desired ``zero" 
pseudo-detector channel:
\begin{equation}
\widetilde z_H(f)=
-\left(\widetilde s_{H_1}(f) - \widetilde s_{H_2}(f) \right)\,
\sqrt{1-\widetilde\alpha(f)-\widetilde\alpha^*(f)+2|\widetilde \alpha(f)|^2}\,,
\label{e:z_H(f)}
\end{equation}
where $\widetilde \alpha(f)$ is given as before (c.f.\ Eq.~(\ref{eqn:alpha1})).
The power spectrum $P_{z}(f)$ of $\widetilde z_H(f)$ is given by 
the eigenvalue $\lambda_2$ above.

\end{widetext}

\begin{figure*}[htbp!]
\includegraphics[width=6.0in,angle=0]{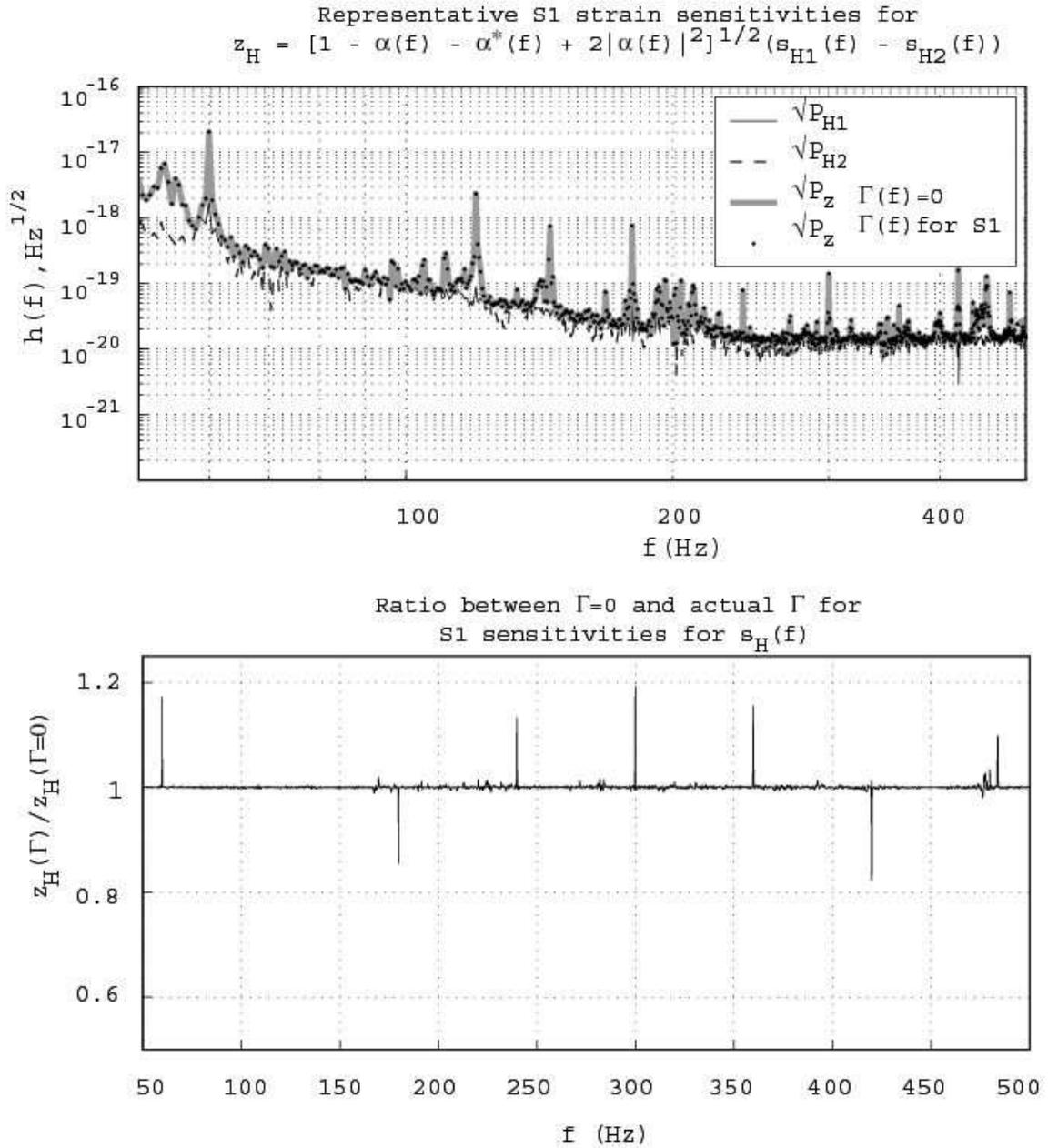}
\caption{Same as Fig.~\ref{fig:Optimal_sH}, but for the
null signal $\widetilde z_H(f)$ instead of the optimal
estimate $\widetilde s_H(f)$. 
Strain spectral densities (i.e., absolute value) of 
$\widetilde z_H(f)$ (gray or dotted), $\widetilde s_{H_1}(f)$ (black), and 
$\widetilde s_{H_2}(f)$ (dashed), representative of the S1 data.
{\bf{Top Panel:}} overlay of the individual amplitude spectral 
densities with that of 
the strain spectral density $|\widetilde z_H(f)|$ is 
calculated with the S1 run-averaged coherence,  $\Gamma_{H_1 H_2}(f)$. 
On this scale, the left hand panel shows no 
discernible difference between the spectra for $\Gamma_{H_1 H_2}(f)$, 
and with $\Gamma_{H_1H_2} = 0$, 
suggesting that even the level of coherence seen during the S1 run 
might be sufficiently low to allow one to simply
combine the L1-H1 and L2-H2 cross-correlation 
measurements under the assumption of zero cross-correlated
noise. 
The optimality of the estimate $\widetilde z_H(f)$ is visible here 
because it is 
{\em{always less than the larger}} of $\widetilde s_{H_1}(f)$  or 
$\widetilde s_{H_2}(f)$. 
{\bf{Bottom panel:}} overlay of individual amplitude spectra with that 
for $|\widetilde z_H(f)|$ 
calculated with $\Gamma_{H_1 H_2}(f)=0$ (i.e., assuming no coherence).  
The difference between the two is very small except for the very 
lowest frequencies and at narrow line features.}
\label{fig:Optimal_zH}
\end{figure*}

Figure~\ref{fig:Optimal_zH} shows plots of the strain spectral 
densities for $\widetilde z_H(f)$, 
$\widetilde s_{H_1}(f)$, and $\widetilde s_{H_2}(f)$, 
representative of the S1 data, similar to Fig.~\ref{fig:Optimal_sH}. 
%

\subsection{Limiting case for zero cross-correlated noise}

In the limit that the two detectors are uncorrelated
(i.e., $\rho_{H_1H_2}(f)=0$), the expression for 
$\widetilde \alpha(f)$ simplifies considerably 
(c.f.~Eq.~(\ref{eqn:alphauncorr})).
In this limit, $\widetilde z_H(f)$ and $P_z(f)$ become
\begin{eqnarray}
\widetilde z_H(f)
&=&
-\left(\widetilde s_{H_1}(f) - \widetilde s_{H_2}(f)\right) 
\frac{\sqrt{P^2_{H_1}(f)+P^2_{H_2}(f)}}
{P_{H_1}(f)+P_{H_2}(f)}\,,
\nonumber\\
&&
\label{eqn:uncorrZh}
\\
P_{z}(f)
&=&
\frac{P^2_{H_1}(f)+P^2_{H_2}(f)}{P_{H_1}(f)+P_{H_2}(f)}\,.
\label{eqn:uncorrPz}
\end{eqnarray}
In particular, note that $P_z(f)$ satisfies the inequality
\begin{eqnarray}
\max\{P_{H_1}(f), P_{H_2}(f)\} - 
\min\{P_{H_1}(f), P_{H_2}(f)\}
\nonumber\\
\le P_{z}(f) \le 
\max\{P_{H_1}(f), P_{H_2}(f)\}\,. 
\end{eqnarray}
This last equation shows that the null channel 
$\widetilde z_H(f)$ contains {\em less\,} noise power than 
the difference of $\widetilde n_{H_1}(f)$, $\widetilde n_{H_2}(f)$. 
The filtering produced by $\widetilde\alpha(f)$ results in a 
less noisy null estimator than the quantity
$\widetilde n_{H_1}(f) - \widetilde n_{H_2}(f)$. 
In the limit that either signal dominates the noise power
(e.g., $P_{H_1}(f) \ll P_{H_2}(f)$), 
\begin{equation}
P_{z}(f) \rightarrow 
\max\{P_{H_1}(f), P_{H_2}(f)\} - 
\min\{P_{H_1}(f), P_{H_2}(f)\}\,.
\end{equation}

In addition, one can form the quantity:
\ba
t(f):=\frac{\widetilde z_{H}(f)}{\sqrt{P_z(f)}} 
= \frac{-\left(\widetilde s_{H_1}(f) - \widetilde s_{H_2}(f)\right) }
{\sqrt{P_{H_1}(f)+P_{H_2}(f)}}.
\ea
As suggested by the label $t$, this quantity is identical to the Student's 
$t$ statistic, which is used to assess the statistical significance of 
two quantities having different means and variances.

\section{Cross-correlation statistics using composite 
pseudo-detector channels}\label{sec:V}
\label{sec:ccstatistic}

Since the instrumental transcontinental (L1-H1, L1-H2) 
cross-correlations are assumed to be negligible, the derivation of 
the optimal filter when using the pseudo-detector channels for 
Hanford proceeds exactly as has been presented in the literature 
\cite{flan, leshouches, allenromano} with $P_{H_1}(f)$, 
$P_{H_2}(f)$ replaced by $P_{H}(f)$, $P_{z}(f)$ for the optimal 
estimate and the null signal, respectively. 

\subsection{Cross-correlation statistic for the optimal estimate
of the gravitational wave strain}

The cross-correlation statistic is given by
\begin{equation}
Y_{L_1 H}:= 
\int_{-T/2}^{T/2} dt_1\int_{-T/2}^{T/2} dt_2\,
s_{L_1}(t_1)\, Q_{L_1 H}(t_1-t_2)\, s_H(t_2)\,,
\label{e:Y_continuous_time}
\end{equation}
where $T$ is the observation time and $Q_{L_1 H}(t)$ is the 
{\em optimal\,} filter, which is chosen to maximize the 
signal-to-noise ratio of $Y_{L_1 H}$.
The corresponding frequency domain expression is
\begin{equation}%
Y_{L_1 H}
\propto
\int_{-\infty}^{\infty} df\,  
\widetilde s_{L_1}^*(f)\,\widetilde Q_{L_1 H}(f)\,\widetilde s_{H}(f)\,.
\label{e:Y_continuous_freq}
\end{equation}
Specializing to the case $\Omega_{\rm gw}(f)\equiv\Omega_0={\rm const}$, 
the optimal filter becomes
\begin{equation}
\widetilde Q_{L_1 H}(f) = 
{\cal N}_{L_1 H}
\frac{\gamma(|f|)}{|f|^3 P_{L_1}(f)P_H(f)}\,,
\label{e:optimal_continuous}
\end{equation}
where ${\cal N}_{L_1 H}$ is a (real) overall normalization constant. 
In practice we choose ${\cal N}_{L_1 H}$ so that the expected
value of the cross-correlation is $\Omega_0\,h_{100}^2\,T$,
where $h_{100}$ is the Hubble expansion rate $H_0$ in units of
$H_{100}:= 100~{\rm km}\,\,{\rm s}^{-1}\,{\rm Mpc}^{-1}$.
For such a choice,
\begin{equation}
{\cal N}_{L_1 H} = 
\frac{20\pi^2}{3 H_{100}^2}\
\left[\int_{-\infty}^\infty df\, 
\frac{\gamma^2(|f|)}{f^6 P_{L_1}(f)P_{H}(f)}
\right]^{-1}\,.
\label{e:normalization_continuous}
\end{equation}
Moreover, one can show that the normalization factor 
${\cal N}_{L_1 H}$ and theoretical variance, 
$\sigma^2_{Y_{L_1 H}}$, of $Y_{L_1 H}$ are related by a 
simple numerical factor:
\begin{equation}
{\cal N}_{L_1 H} = 
\frac{1}{T}
\left(\frac{3H^2_{100}}{5\pi^2}\right)
\sigma^2_{Y_{L_1 H}}\,.
\label{e:norm_sigma_reln}
\end{equation}
%

\subsubsection{Limiting case for white coherence and 
$P_{H_1}(f)\propto P_{H_2}(f)$}
\label{sec:seca1}

If the coherence is white (i.e., $\rho_{H_1 H_2}(f)={\rm const}$)
and the power spectra $P_{H_1}(f)$, $P_{H_2}(f)$ are proportional 
to one another, then one can show that the value of the 
cross-correlation statistic $Y_{L_1 H}$ 
reduces to a linear combination of the cross-correlation statistics
$Y_{L_1 H_1}$ and 
$Y_{L_1 H_2}$ calculated separately for L1-H1 and L1-H2,
if we allow for instrumental correlations between H1 and H2.
Thus, for this case, combining the point estimates of $\Omega_0$
made separately for L1-H1 and L1-H2 gives the same result as 
performing the coherent pseudo-detector channel analysis using the 
single optimal estimator $\widetilde s_H(f)$. 

To show that this is indeed the case, note that 
$\rho_{H_1 H_2}(f)={\rm const}$ implies 
\begin{equation}
\Gamma_{H_1 H_2}(f):=
\abs{\rho_{H_1 H_2}(f)}^2=
{\rm const}\,.\label{eq:gammaconst}
\end{equation}
We will drop subscripts for constant quantities. If we further assume that $P_{H_2}(f)= \eta  P_{H_1}(f)$, then 
\begin{equation}
\frac{P_{H_1 H_2}(f)}{P_{H_2}(f)}=\frac{\rho}{\sqrt{\eta}}\,,
\quad
\frac{P^*_{H_1 H_2}(f)}{P_{H_1}(f)}=\rho^*\sqrt{\eta}\,.\label{eq:rhoconst}
\end{equation}
Thus, the integrand of the cross-correlation statistic, 
\begin{equation}
Y_{L_1 H}(f) :=
\widetilde s^*_{L_1}(f)\,\widetilde Q_{L_1 H}(f)\,\widetilde s_{H}(f) \,,
\end{equation}
becomes
\bw
\begin{eqnarray}
\frac{Y_{L_1H}(f)}{{\cal N}_{L_1 H}}
&=&
\frac{\gamma(f) \widetilde s^*_{L_1}(f)
\left[
\widetilde s_{H_1}(f) (P_{H_2}(f) - P_{H_1 H_2}(f))+
\widetilde s_{H_2}(f) (P_{H_1}(f) - P^*_{H_1 H_2}(f))
\right]}
{|f|^3 P_{L_1}(f) P_{H_1}(f) P_{H_2}(f)\,(1-\Gamma_{H_1 H_2}(f))}
\\
&=&
\frac{1}{1-\Gamma}\,
\left[
\left(1-\frac{\rho}{\sqrt{\eta}}\right)\, 
\frac{Y_{L_1 H_1}(f)}{{\cal N}_{L_1 H_1}}+
\left(1-\rho^*\sqrt{\eta}\right)\,
\frac{Y_{L_1 H_2}(f)}{{\cal N}_{L_1 H_2}} 
\right]
\,,
\label{e:Y_L1H(f)}
\end{eqnarray}
where the normalization factor 
Eq.~(\ref{e:normalization_continuous}) is
\begin{eqnarray}
{\cal N}_{L_1 H}
&=&
\frac{20 \pi^2}{3 H_{100}^2}
\left[
\int_{-\infty}^\infty df\,
\frac{\gamma^2(|f|)
(P_{H_1}(f)+P_{H_2}(f)-(P_{H_1 H_2}(f)+P^*_{H_1 H_2}(f)))}
{f^6 P_{L_1}(f) P_{H_1}(f) P_{H_2}(f)\,(1-\Gamma_{H_1 H_2}(f))}
\right]^{-1}
\\
&=&
(1-\Gamma)\,
\left[
\left(1-\frac{\rho}{\sqrt{\eta}}\right)\, 
{\cal N}_{L_1 H_1}^{-1} + 
\left(1-\rho^*\sqrt{\eta}\right)\,
{\cal N}_{L_1 H_2}^{-1}
\right]^{-1}\,.
\end{eqnarray}
Equivalently,
\begin{eqnarray}
\sigma_{Y_{L_1 H}}^2
&=&
(1-\Gamma)\,
\left[
\left(1-\frac{\rho}{\sqrt{\eta}}\right)\, 
\sigma^{-2}_{Y_{L_1 H_1}}+
\left(1-\rho^*\sqrt{\eta}\right)\, 
\sigma^{-2}_{Y_{L_1 H_2}}
\right]^{-1}
\\
&=&
(1-\Gamma)\,
\frac{\sigma^2_{Y_{L_1 H_1}}\sigma^2_{Y_{L_1 H_2}}}
{\sigma^2_{Y_{L_1 H_1}}
\left(1-\rho^*\sqrt{\eta}\right)+  
\sigma^2_{Y_{L_1 H_2}}
\left(1-\frac{\rho}{\sqrt{\eta}}\right)\,,}
\label{e:sigma_L1H^2}
\end{eqnarray}
where we used 
Eq.~(\ref{e:norm_sigma_reln}) and similar equations
to relate ${\cal N}_{L_1 H_1}$, ${\cal N}_{L_1 H_2}$ to 
$\sigma^2_{L_1 H_1}$, $\sigma^2_{L_1 H_2}$.

Substituting the above results for the normalization factors and 
variances into Eq.~(\ref{e:Y_L1H(f)}) and integrating over frequency, 
we find:
\begin{eqnarray}
Y_{L_1 H}
&=&
\frac{\sigma_{Y_{L_1 H}}^2}{(1-\Gamma)}\,
\left[
\left(1-\frac{\rho}{\sqrt{\eta}}\right)\, 
\frac{Y_{L_1 H_1}}{\sigma^2_{Y_{L_1 H_1}}}+
\left(1-\rho^*\sqrt{\eta}\right)\,
\frac{Y_{L_1 H_2}}{\sigma^2_{Y_{L_1 H_2}}} 
\right]
\\
&=&
\frac{\sigma^2_{Y_{L_1 H_1}}\sigma^2_{Y_{L_1 H_2}}}
{\sigma^2_{Y_{L_1 H_1}}
\left(1-\rho^*\sqrt{\eta}\right)+  
\sigma^2_{Y_{L_1 H_2}}
\left(1-\frac{\rho}{\sqrt{\eta}}\right)}\,
\left[
\left(1-\frac{\rho}{\sqrt{\eta}}\right)\, 
\frac{Y_{L_1 H_1}}{\sigma^2_{Y_{L_1 H_1}}}+
\left(1-\rho^*\sqrt{\eta}\right)\,
\frac{Y_{L_1 H_2}}{\sigma^2_{Y_{L_1 H_2}}} 
\right]
\\
&=&
\frac
{\sigma_{Y_{L_1 H_2}}^2\,
\left(1-\frac{\rho}{\sqrt{\eta}}\right)\,
Y_{L_1 H_1} +
 \sigma_{Y_{L_1 H_1}}^2\,
\left(1-\rho^*\sqrt{\eta}\right)\,
Y_{L_1 H_2}}
{\sigma^2_{Y_{L_1 H_1}}
\left(1-\rho^*\sqrt{\eta}\right)+  
\sigma^2_{Y_{L_1 H_2}}
\left(1-\frac{\rho}{\sqrt{\eta}}\right)}\,.
\end{eqnarray}
\end{widetext}
Or in the notation of Appendix~\ref{sec:general_method}:
\begin{equation}
Y_{L_1 H}
=
\frac
{(C_{22} - C_{12})\,Y_1 + (C_{11} - C_{21})\,Y_2}
{C_{11}+C_{22}-C_{12}-C_{21}}\,,
\end{equation}
where $Y_1:=Y_{L_1 H_1}$, $Y_2:=Y_{L_1 H_2}$, and where we 
used Eqs.~(\ref{e:C11C22}), (\ref{e:limiting_C12C21}) 
from Appendix~\ref{sec:application} to equate
$\sigma_{Y_{L_1 H_1}}^2$, $\sigma_{Y_{L_1 H_2}}^2$ 
with $C_{11}$, $C_{22}$, and
$P_{H_1 H_2}/P_{H_2}\equiv \rho/\sqrt{\eta}$,
$P^*_{H_1 H_2}/P_{H_1}\equiv \rho^* \sqrt{\eta}$ 
with $C_{12}/C_{22}$, $C_{21}/C_{11}$.
Thus, we see that for the limiting case of white coherence and 
proportional power spectra, the pseudo-detector optimal estimator 
analysis reduces to a relatively simple combination of the 
separate cross-correlation statistic measurements.

Finally, note that in the case of zero cross-correlated noise
(i.e., for $\rho_{H_1 H_2}(f)=0$) we get
\begin{eqnarray}
Y_{L_1 H}
&=&
\frac
{\sigma_{Y_{L_1 H_2}}^2\,Y_{L_1 H_1} +
 \sigma_{Y_{L_1 H_1}}^2\,Y_{L_1 H_2}}
{\sigma^2_{Y_{L_1 H_1}}+\sigma^2_{Y_{L_1 H_2}}}
\\
&=&
\frac
{\sigma_{Y_{L_1 H_1}}^{-2}\,Y_{L_1 H_1}+
 \sigma_{Y_{L_1 H_2}}^{-2}\,Y_{L_1 H_2}}
{\sigma_{Y_{L_1 H_1}}^{-2}+\sigma_{Y_{L_1 H_2}}^{-2}}\,,
\label{e:no_correlations}
\end{eqnarray}
which is the standard method of combining results of measurements
in the absence of correlations \cite{S1_stochpaper}.

\subsection{Cross-correlation statistic for the null signal}

Once again, the cross-correlation statistic in the frequency 
domain is given by
\begin{equation}
Y_{L_1 z}\propto
\int_{-\infty}^{\infty}df\,
\widetilde s^*_{L_1}(f)\,\widetilde Q_{L_1 z}(f)\,\widetilde z_{H}(f)\,.
\label{e:Y_continuous_freq1}
\end{equation}
As before, the optimal filter for 
$\Omega_{\rm gw}(f)\equiv\Omega_0={\rm const}$ is
\begin{equation}
\widetilde Q_{L_1 z}(f) = 
{\cal N}_{L_1 z} 
\frac{\gamma(|f|)}{|f|^3 P_{L_1}(f)P_z(f)}\,,
\label{e:optimal_continuous2}
\end{equation}
where ${\cal N}_{L_1 z}$ is chosen to be
\begin{eqnarray}
{\cal N}_{L_1 z} &=& 
\frac{20\pi^2}{3 H_{100}^2}\
\left[\int_{-\infty}^\infty df\, 
\frac{\gamma^2(|f|)}{f^6 P_{L_1}(f)P_{z}(f)}
\right]^{-1}
\label{e:normalization_continuous1}
\end{eqnarray}
and is related to the theoretical variance $\sigma^2_{Y_{L_1 z}}$ via:
\begin{equation}
{\cal N}_{L_1 z} = 
\frac{1}{T}
\left(\frac{3 H_{100}^2}{5\pi^2}\right)
\sigma^2_{Y_{L_1 z}}\,.
\end{equation}
%

\subsubsection{Limiting case for white coherence and 
$P_{H_1}(f)\propto P_{H_2}(f)$}

We start again with the same assumptions that the coherence is white
and the power spectra $P_{H_1}(f)$, $P_{H_2}(f)$ are proportional 
to one another (cf.~Eqs.~(\ref{eq:gammaconst}), (\ref{eq:rhoconst})).
Then it is possible to show that the value of the 
cross-correlation statistic $Y_{L_1 z}$ 
reduces to a linear combination of the cross-correlation statistics
$ Y_{L_1 H_1}$ and 
$Y_{L_1 H_2}$ calculated separately for L1-H1 and L1-H2,
if we allow for instrumental correlations between H1 and H2.
After much algebra similar to that presented earlier in Section \ref{sec:seca1} we obtain: 
\begin{widetext}
\begin{equation}
\frac{Y_{L_1z}}{\sigma^2_{Y_{L_1z}}} =
\frac{{\sqrt{\eta \,\left( {\eta }^{\frac{3}{2}} - 
          \rho  \right) }}\,
    \left(  \eta  \, \frac{Y_{L_1H_2}}
       {{\sigma^2_{Y_{L_1H_2}}}}    - \frac{Y_{L_1H_1}}
         {{\sigma^2_{Y_{L_1H_1}}}} \right) }{\sqrt{\left( {\eta }^{\frac{3}{2}} - \rho^* \right) \,
       \left( \eta  + {\eta }^3 + 
         2\,|\rho |^2 - 
         \left(\sqrt{\eta }+{\eta }^{\frac{3}{2}} \right)\,
          \left( \rho  + \rho^* \right)  
         \right) }}\,,
\end{equation}
or, equivalently,
\ba
\frac{Y_{L_1z}}{\sigma_{Y_{L_1z}}} &=&
    {\sqrt{\frac{{\eta }^{\frac{3}{2}} - \rho }
      {\left( {\eta }^{\frac{3}{2}} - 
          \rho^* \right) \,
        \left( \eta  + {\eta }^2 - 
          {\sqrt{\eta }}\,
           \left( \rho  + \rho^* \right) 
          \right) }}}\, \left(  \eta  \, \frac{Y_{L_1H_2}}{\sigma_{Y_{L_1H_2}}} 
               - \sqrt{\eta} \frac{Y_{L_1H_1}}
         {\sigma_{Y_{L_1H_1}}}
          \right) \\
&=&
    {\sqrt{\frac{{\eta }^{\frac{3}{2}} - \rho }
      {\left( {\eta }^{\frac{3}{2}} - 
          \rho^* \right) \,
        \left( 1 - 
          \frac{{\sqrt{\eta }}\,
           \left( \rho  + \rho^* \right) }{\eta  + {\eta }^2}
          \right) }}}\, \left(   \frac{Y_{L_1H_2} -Y_{L_1H_1}}{\sqrt{\sigma^2_{Y_{L_1H_2}}+\sigma^2_{Y_{L_1H_1}}}}          \right)\,.
              \ea
\end{widetext}

\subsubsection{Limiting case for zero cross-correlated noise}

If also $\rho_{H_1H_2}(f)=0$, then the two interferometer noise floors 
are uncorrelated, and the cross-correlation statistic $Y_{L_1 z}$ 
for the null channel simplifies further:
\ba
\frac{Y_{L_1 z}}{\sigma_{L_1z}} &=& \frac{Y_{L_1 H_2} -Y_{L_1 H_1}}{\sqrt{\sigma^2_{Y_{L_1H_1}}+ \sigma^2_{Y_{L_1H_2}}}}\label{eq:studentt}
\ea
Equation~(\ref{eq:studentt}) shows that in this limit the quantity 
$Y_{L_1 z}/\sigma_{L_1z}$ follows the Student's $t$ distribution. This distribution provides a measure to assess the significance of the difference between two experimental quantities having different means and variances. Here it provides a measure of consistency of the two independent measurements, $Y_{L_1H_1}$and $Y_{L_1H_2}$: their difference should be consistent with zero within the combined experimental errors.

\subsection{Combining triple and double coincident measurements}

In order to make use of this method for the analysis of 
future science data, we will need to partition the data into 
three {\em non-overlapping\,} (hence statistically independent) 
sets: the H1-H2-L1 triple coincident data set, and the two 
L1-H1 and L1-H2 double coincident data sets. 
The triple coincidence data would be analyzed in the manner 
described in this paper, while the double coincidence data 
(corresponding to measurements from different epochs or from
different science runs) can be simply combined under the 
assumption of statistical independence 
(cf.\ Eq.~(\ref{e:no_correlations})).

\section{Conclusion}
\label{sec:VI}

The approach presented above is fundamentally different from 
how the analysis of S1 data was conducted and represents a manner
to maximally exploit the feature of LIGO that has two co-located 
interferometers. 
This technique is possible for the Hanford pair of detectors 
because, to high accuracy, the gravitational wave signature is 
guaranteed to be {\em identically\,} imprinted on both data streams. 
Coherent, time-domain mixing of the two interferometer strain 
channels can thus be used to optimal advantage to
provide the best possible estimate of the gravitational wave strain, 
and to provide a null channel with which any gravitational wave 
analysis can be calibrated for backgrounds.

An analogous technique of ``time-delay interferometry" (TDI) has been proposed in the context of the Laser Interferometer Space Array ({\em{LISA}}) concept \cite{LISA1} \cite{LISA2}.  However, in that case the data analysis is very different from what is explored in our paper. TDI involves time-shifting the 6 data-streams of LISA (2 per
arm) appropriately before combining them so as to cancel (exactly) the
laser-frequency noise that dominates other LISA noise sources. Even after implementing TDI, the resulting data combinations (with the
laser-frequency noise eliminated) are not all independent, and may have
cross-correlated noises from other, non-gravitational-wave, sources. One, therefore, seeks in
LISA data analysis an optimal strategy for detecting a given signal in
these TDI data combinations. On the other hand, the method presented in this paper is not about canceling specific noise components from data; rather, it is about deducing the optimal detection strategy in the presence of cross-correlated noise.

The usefulness of $\widetilde z_H(f)$ is that it may be used to 
analyze cross-correlations for non-gravitational wave signals 
between the Livingston and Hanford sites. 
This would enable a {\em null\,} measurement to be made---i.e., 
one in which gravitational radiation had been effectively 
``turned off." 
In this sense, using $\widetilde z_H(f)$ would be analogous to 
analyzing the ALLEGRO-L1 correlation when the orientation of the 
cryogenic resonant bar detector ALLEGRO is at $45^\circ$ with 
respect to the interferometer arms \cite{lazzarinifinn,allegrocqg}. 
Under suitable analysis, the cross-correlation statistic 
$Y_{L_1 z}$ could be used to establish an ``off-source" 
background measurement for the stochastic gravitational wave 
background. 

Ultimately, the usefulness of such a null test will be related 
to how well the relative calibrations between H1 and H2 are known. 
If the contributions of $\widetilde h(f)$ to $\widetilde s_{H_1}(f)$ 
and $\widetilde s_{H_2}(f)$ are not equal due to calibration 
uncertainties, then this error will propagate into the generation 
of $\widetilde s_H(f)$, $\widetilde z_H(f)$. 
It is possible to estimate this effect as follows. 
Due to the intended use of $\widetilde z_H(f)$ in a null measurement, 
the leakage of $\widetilde h(f)$ into this channel is the greater 
concern. 
Considering the structure of Eqs.~(\ref{eqn:defSh}), (\ref{eqn:defZh1}), 
(\ref{eqn:defZh2}), it is clear that effects of {\em differential\,} 
calibration errors in $\widetilde s_H(f)$ will tend to average out, 
whereas such errors will be {\em amplified\,} in  $\widetilde z_H(f)$. 
Assume a differential calibration error of 
$\pm \widetilde \epsilon(f)$. 
Then $\widetilde z_H(f)$ will contain a gravitational wave signature
\begin{equation}
\delta \widetilde h(f) = 2\widetilde \epsilon(f) \widetilde h(f)\,,
\end{equation}
with corresponding power
\begin{equation}
\delta P_h(f) = 4|\widetilde \epsilon(f)|^2 P_h(f)\,.
\end{equation}
The amplitude leakage affects single-interferometer based analyses; 
the power leakage affects multiple interferometer correlations 
(such as the stochastic background search). 
Assuming reasonably small values for $\pm \widetilde \epsilon(f)$, 
if a search sets a threshold $\rho_*$ on putative gravitational 
wave events detected in channel $\widetilde s_H(f)$, then the 
corresponding contribution in $\widetilde z_H(f)$ would be 
approximately $2 |\overline \epsilon| \rho_*$, where 
$|\overline \epsilon|$ denotes the magnitude of the frequency 
integrated differential calibration errors. 
For any reasonable threshold (e.g., $\rho_*\approx 10$) above 
which one would claim a detection, and for typical differential 
calibration uncertainties of $2|\overline \epsilon|\alt 20\%$, 
then the same event would have a signal-to-noise level of 
$\rho_*\approx 2$ in the null channel, well below what one would 
consider meaningful. 
A more careful analysis is needed to quantify these results, since 
calibration uncertainties also propagate into $\widetilde \alpha(f)$. 

While the focus of this paper is the application of this technique 
to the search for stochastic gravitational waves, it appears 
that {\em any\,} analysis can exploit this approach. 
It should be straightforward to tune pipeline filters and cull 
spurious events by using the null channel to veto events seen in 
the $\widetilde s_H(f)$ channel.

\begin{acknowledgments}

One of the authors (AL) wishes to thank Sanjeev Dhurandhar for 
his hospitality at IUCAA during which the paper was completed. 
He provided helpful insights by pointing out the geometrical nature 
of the signals and their inherent three dimensional properties 
that span the space $\{\hat n_{H_1},\hat n_{H_2}, \hat h\}$. 
This led to an understanding of how diagonalization of the 
covariance matrix could be achieved only after properly removing 
the signature of $h$ from the interferometer signals. 
The authors gratefully acknowledge the careful review and helpful 
suggestions provided by Nelson Christensen which helped finalize 
the manuscript.

This work was performed under partial funding from the following NSF Grants:
PHY-0107417, 0140369, 0239735, 0244902, 0300609, and INT-0138459. 
JDR and TR acknowledge partial support on PPARC Grant PPA/G/O/2001/00485.
This document has been assigned LIGO Laboratory document number 
\DCCNum.
\end{acknowledgments}


\appendix

\section{General method of combining measurements allowing
for cross-correlations}\label{sec:general_method}

In this appendix, we present a general method of combining 
measurements, allowing for possible correlations between them.
In the following appendix (Appendix~\ref{sec:application}), 
this method is applied to the case 
of the L1-H1 and L1-H2 cross-correlation statistic measurements, 
which are taken over the same observation period and which 
may contain significant instrumental H1-H2 correlations.

It is important to emphasize that the method discussed 
in this appendix is {\em not\,} the same as the pseudo-detector 
optimal estimator method discussed in the main text; 
the pseudo-detector method combines the 
data at the level of data streams $\widetilde s_{H_1}(f)$,
$\widetilde s_{H_2}(f)$ {\em before\,} optimal filtering, 
while the method discussed here combines the data at the level 
of the cross-correlation statistic measurements 
$Y_{L_1 H_1}$ and $Y_{L_1 H_2}$---i.e., 
{\em after\,} optimal filtering of the individual data streams.
As such, the method described here is not optimal, in 
general, since it does {\em not\,} take advantage of the common 
gravitational wave strain component $h$ present in H1 and H2.
However, as shown in the main text, when the cross-correlation
$\rho_{H_1 H_2}(f)$ is white and the power spectra $P_{H_1}(f)$,
$P_{H_2}(f)$ are proportional to one another, the pseudo-detector 
optimal estimator method reduces to the method 
described here.

Consider then a pair of (real-valued) random variables $Y_1$, 
$Y_2$ with the same theoretical mean
\begin{equation}
\mu:=\langle Y_1\rangle =\langle Y_2\rangle\,,
\end{equation}
and covariance matrix
\begin{equation}
||{\bf C}||
:=
\left[
\begin{array}{cc}
C_{11} & C_{12} \\
C_{21} & C_{22}
\end{array}
\right]\,,
\end{equation}
where
\begin{equation}
C_{ij}:=
\langle(Y_i-\mu)(Y_j-\mu)\rangle= 
\langle Y_i Y_j\rangle - \mu^2\,.
\end{equation}
Note that $C_{12}=C_{21}$ since the $Y_i$ are real.
The absence of cross-correlations corresponds to $C_{12}=C_{21}=0$.

Now form the weighted average
\begin{equation}
Y_{\rm opt}:=
\frac{\sum_i \lambda_i Y_i}
{\sum_j \lambda_j}\,.
\end{equation}
It is straightforward to show that $Y_{\rm opt}$ has 
theoretical mean $\mu_{\rm opt}=\mu$, and theoretical variance
\begin{equation}
\sigma_{\rm opt}^{2} = 
\frac{1}{(\sum_k \lambda_k)^2}\,
\sum_i \sum_j \lambda_i C_{ij} \lambda_j\,.
\end{equation}
Now find the weighting factors $\lambda_i$ that minimize
the variance of $Y_{\rm opt}$.
The result is
\begin{equation}
\lambda_i = \sum_j ||{\bf C}||^{-1}_{ij}\,,
\label{e:lambda_i}
\end{equation}
or, explicitly,
\begin{equation}
\lambda_1 = 
\frac{C_{22}-C_{12}}{\det||{\bf C}||}\,,
\quad
\lambda_2 = 
\frac{C_{11}-C_{21}}{\det||{\bf C}||}\,,
\end{equation}
where $\det||{\bf C}||:=C_{11}C_{22}-C_{12}C_{21}$.

One can prove the above result by defining an inner product
\begin{equation}
({\bf A},{\bf B}):=\sum_i\sum_j A_i ||{\bf C}||^{-1}_{ij} B_j\,,
\end{equation}
and rewriting the variance as
\begin{equation}
\sigma_{\rm opt}^2 = 
\frac{({\bf C}\cdot {\bf \lambda},{\bf C}\cdot {\bf \lambda})}
{({\bf C}\cdot {\bf \lambda}, 1)^2}\,.
\end{equation}
Then $\sigma_{\rm opt}^2$ is minimized by choosing $\lambda_i$ 
such that 
\begin{equation}
{\bf C}\cdot{\bf \lambda}:=\sum_j C_{ij}\lambda_j = 1
\end{equation}
for all $i$.

For such a choice,
\begin{eqnarray}
\sigma^{-2}_{\rm opt}
&=&
\sum_i \lambda_i=
\frac{C_{11}+C_{22}-C_{12}-C_{21}}{\det||{\bf C}||}\,,
\\
\frac{Y_{\rm opt}}{\sigma_{\rm opt}^2}
&=&
\frac
{(C_{22}-C_{12})\, Y_1 + (C_{11}-C_{21})\, Y_2}
{\det||{\bf C}||}\,,
\end{eqnarray}
so
\begin{equation}
Y_{\rm opt} =
\frac
{(C_{22}-C_{12})\, Y_1 + (C_{11}-C_{21})\, Y_2}
{C_{11}+C_{22}-C_{12}-C_{21}}\,.
\label{e:Yopt}
\end{equation}
This is the desired combination.

\section{Application of the general method to the L1-H1, L1-H2 
cross-correlation statistic measurements}\label{sec:application}

Here we apply the results of the previous appendix to the 
L1-H1 and L1-H2 cross-correlation measurements.
We let $Y_1$ denote the cross-correlation statistic 
$Y_{L_1 H_1}$ for the L1-H1 detector pair, and $Y_2$ denote 
the cross-correlation statistic $Y_{L_1 H_2}$ for L1-H2, and
assume that the measurements are taken over the {\em same\,} 
observation period of duration $T$.
(If the observations were over different times, then there
would be no cross-correlation terms and a simple weighted 
average by $\sigma_i^{-2}$ would suffice.)
We need only calculate the components of the covariance 
matrix to apply the method described in the previous appendix.

To calculate the $C_{ij}$, we assume (as in the main text)
that the cross-correlated stochastic signal power 
$P_\Omega(f)$ is small compared to the auto-correlated noise 
in the individual detectors, and that there are no broadband 
transcontinental instrumental or environmental correlations---i.e., 
$|P^n_{L_1 H_i}(f)|$ is small compared to the 
auto-correlated noise, the cross-correlated stochastic signal
power, and the H1-H2 cross-correlation $|P_{H_1 H_2}(f)|$.
Then it is fairly straightforward to show that
\begin{equation}
C_{11}=\sigma^2_{L_1 H_1}\,,
\quad
C_{22}=\sigma^2_{L_1 H_2}\,,
\label{e:C11C22}
\end{equation}
and
\begin{eqnarray}
&&
\frac{C_{12}}{C_{11} C_{22}}=
\frac{C_{21}}{C_{11} C_{22}}
\nonumber\\
&&\quad
=\frac{1}{T}
\left(
\frac{3 H_{100}^2}{10\pi^2}
\right)^2
\int_{-\infty}^\infty df\,
\frac{\gamma^2(|f|)P_{H_1 H_2}(f)}
{f^6 P_{L_1}(f) P_{H_1}(f) P_{H_2}(f)}\,.
\nonumber\\
\end{eqnarray}
Note that the above integral is real since 
$P_{H_1 H_2}(-f)=P^*_{H_1 H_2}(f)$ and the integration is over
all frequencies (both positive and negative).

If we further consider the limiting case defined by white 
coherence (i.e., $\rho_{H_1 H_2}(f)={\rm const}$) and 
proportional power spectra (i.e., $P_{H_1}(f)\propto P_{H_2}(f)$), 
then 
$P_{H_1 H_2}(f)/P_{H_1}(f)$ and $P^*_{H_1 H_2}(f)/P_{H_2}(f)$ 
are both constant with values
\begin{equation}
\frac{P_{H_1 H_2}}{P_{H_2}}=
\frac{C_{12}}{C_{22}}\,,
\quad
\frac{P^*_{H_1 H_2}}{P_{H_1}}=
\frac{C_{21}}{C_{11}}\,.
\label{e:limiting_C12C21}
\end{equation}

\end{document}